\documentclass{aa}
\usepackage{graphics}
\newcommand\jcd{Christensen-Dalsgaard}
\newlength{\figwidth}
\begin{document}
\thesaurus{06.01.1; 06.09.1; 06.15.1}
\title{Limits on the proton-proton reaction
cross-section from helioseismology}
\titlerunning{Limits on pp reaction cross-section from helioseismology}
\author{H. M. Antia \and S. M. Chitre}
\authorrunning{Antia \and Chitre}
\offprints{H. M. Antia}
\institute{Tata Institute of Fundamental Research,
Homi Bhabha Road, Mumbai 400005, India\\
email: antia@tifr.res.in, chitre@tifr.res.in}
\date{Received }
\maketitle

\begin{abstract}
Primary inversions of  solar oscillation frequencies coupled
with the equations of thermal equilibrium and other input physics,
enable us to
infer the temperature and hydrogen abundance profiles inside the Sun.
These profiles also help in setting constraints on the input physics
that is consistent with the accurately measured oscillation
frequencies data. Helioseismic limits
on the cross-section of proton-proton nuclear reaction as
a function of heavy element abundance in the
solar core are derived. We demonstrate that it is not possible to infer the
heavy element abundance profile, in addition to temperature and hydrogen
abundance profiles, with the helioseismic constraints.

\end{abstract}

\section{Introduction}

The precisely measured frequencies of solar oscillations provide us with a
powerful tool to probe the solar interior with sufficient accuracy.
These frequencies are primarily determined by the mechanical quantities
like sound speed, density or the adiabatic index of the solar material.
The primary inversions of the observed frequencies yield only
the sound speed and density profiles inside the Sun.
On the other hand, in order to infer the temperature and chemical
composition profiles, additional assumptions regarding the input
physics such as opacities, equation of state and nuclear energy
generation rates are required.
Gough \& Kosovichev (\cite{dog88}) and Kosovichev (\cite{kos96})
have employed the equations of thermal equilibrium to express the changes
in primary variables ($\rho,\Gamma_1$) in terms of those in
secondary variables ($Y,Z$) and thus obtained
equations connecting the frequency differences to variations in abundance
profiles.
Shibahashi \& Takata~(\cite{st96}), Takata \& Shibahashi~(\cite{tak98}) and
Shibahashi, Hiremath \& Takata (\cite{shi98}) adopt the equations of
thermal equilibrium, standard
opacities and nuclear reaction rates to deduce the temperature
and hydrogen abundance profiles with the use of only the inverted sound
speed profile. Antia \& Chitre (\cite{ac95}, \cite{ac98})
followed a similar approach, but they used the inverted
density profile, in addition to the sound speed profile, for calculating the
temperature and hydrogen abundance profiles, for a prescribed heavy element
abundance ($Z$) profile.

In general, the computed luminosity in a seismically computed solar
model is not expected to be in agreement with the observed solar luminosity.
By applying
the observed luminosity constraint it is possible to estimate the
cross-section of proton-proton (pp) nuclear reaction.
Antia \& Chitre (\cite{ac98})
estimated this cross-section to be
$S_{11}= (4.15\pm0.25)\times10^{-25}$ MeV barns.
Similar values have been obtained by comparing the computed solar
models with helioseismic data
(Degl'Innocenti, Fiorentini \& Ricci~\cite{inn98};
Schlattl, Bonanno \& Paterno~\cite{bon99}).
The main source of error in these
estimates is the uncertainties in $Z$ profiles.
In this work we try to
find the region in the $Z$--$S_{11}$ plane that is consistent with
the constraints imposed by the helioseismic data.

It may even be argued that one can determine
the pressure, in addition to the sound
speed and density, from primary inversions using the equation of
hydrostatic equilibrium. This profile can then be used as an additional
constraint for determining the heavy element abundance profile.
In this work we  explore the possibility of determining the
$Z$ profile in addition to the $X$ profile
using this additional input. Alternately, we can determine the $Z$
profile (or opacities) instead of the $X$ profile
(Tripathy \& \jcd~\cite{tri98}).
Roxburgh~(\cite{rox96}) has also examined $X$ profiles which are
obtained by suitably scaling the hydrogen abundance profiles from a
standard solar model in order to generate the observed luminosity.
The motivation of this study was to explore the possibility of
reducing the neutrino fluxes yielded by the seismic models by
allowing for variations in both the composition profiles as well as
selected nuclear reaction rates.

\section{The technique}

The sound speed and density profiles inside the Sun are inferred from
the observed frequencies using a Regularized Least Squares technique
(Antia \cite{a96}). The primary inversions based on the equations of
hydrostatic equilibrium along with the adiabatic oscillation equations,
however, give only the mechanical variables like
pressure, density and sound speed. This provides us with the ratio
$T/\mu$, where $\mu$ is the mean molecular weight.
In order to
determine $T$ and $\mu$ separately, it becomes necessary to use the
equations of thermal equilibrium, i.e.,
\begin{eqnarray}
L_r&=&-{64\pi r^2\sigma T^3\over 3\kappa\rho}{dT\over dr},\label{dtr}\\
{dL_r\over dr}&=&4\pi r^2\rho\epsilon,\label{dlr}
\end{eqnarray}
where $L_r$ is the total energy generated within a sphere of radius $r$,
$\sigma$ is the Stefan-Boltzmann constant, $\kappa$ is the Rosseland
mean opacity, $\rho$ is the density and $\epsilon$ is the nuclear energy
generation rate per unit mass. In addition, the equation of state
needs to be adopted to relate the sound speed to chemical composition
and temperature: $c=c(T,\rho,X,Z)$. These three equations are sufficient
to determine the three unknowns $T,L_r,X$, provided the $Z$ profile is
prescribed (Antia \& Chitre \cite{ac98}).

The resulting seismic model will not in general have the correct solar
luminosity which is an observed quantity. It turns out that we need to
adjust the nuclear reaction rates slightly to obtain the correct luminosity
and we believe this boundary condition can be profitably used for
constraining the nuclear reaction rates.
The rate of nuclear energy generation in the Sun is mainly controlled by the
cross-section for the pp nuclear reaction, which has not been measured
in the laboratory.
This nuclear reaction rate is thus calculated theoretically
and it would be interesting to test the validity of calculated
results using the helioseismic constraints. Since the computed
luminosity in seismic models also depends on $Z_c$, the heavy element
abundance in solar core, we attempt to determine the region
in the $Z_c$--$S_{11}$ plane which yields the correct solar luminosity.

Using the density profile along with the equation
of hydrostatic equilibrium, it should be possible to determine the pressure
profile also from primary inversions.
It may even be argued that if we use the
additional constraint, $p=p(T,\rho,X,Z)$ it should be possible to determine
the $Z$ profile besides other profiles. However, it is not clear if these
constraints are independent and in section~3.2
we examine this possibility.

\section{Results}

We use the observed frequencies from GONG (Global Oscillation
Network Group) data for months 4--10
(Hill et al.~\cite{hil96}) which corresponds to the period from
23 August 1995 to 30 April 1996, to calculate the sound speed
and density profiles. A Regularized Least Squares (RLS)
technique for inversion is adopted for this purpose.
With the help of the inverted profiles for sound speed and density,
along with the $Z$ profile from Model~5
of Richard et al.~(\cite{ric96}), we obtain the temperature and hydrogen
abundance profiles by employing the equations of thermal equilibrium.
We adopt the OPAL opacities (Iglesias \& Rogers \cite{igl96}),
OPAL equation of state (Rogers, Swenson \& Iglesias \cite{rog96})
and nuclear reaction rates from
Adelberger et al.~(\cite{fusion}) for obtaining the thermal structure.
Recently, Elliot and Kosovichev~(\cite{ell98}) have
demonstrated that inclusion
of relativistic effects in the equation of state improves the agreement
with helioseismic data. Since the OPAL equation of state does not
include this effect we have applied corrections as outlined by
Elliot and Kosovichev~(\cite{ell98}) to incorporate the relativistic
effects.
The inferred mean molecular weight profile is displayed in Fig.~1.
The only difference between the present calculations and earlier
work of Antia \& Chitre (\cite{ac98}) is in the adopted nuclear reaction rates
and application of the relativistic correction to the equation of
state.

\begin{figure}
\resizebox{\figwidth}{!}{\includegraphics{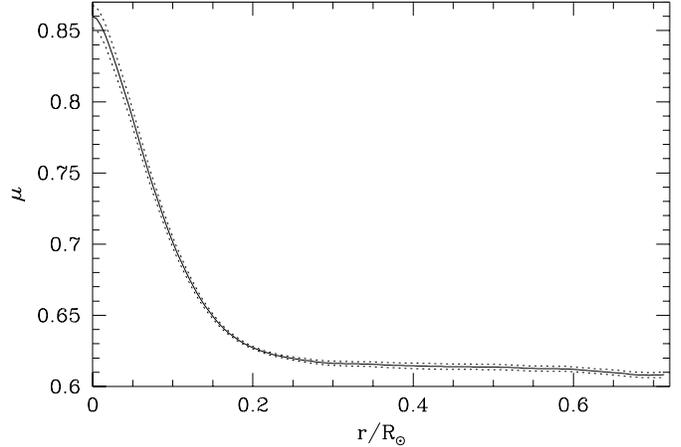}}
\caption{
The mean molecular weight, $\mu$, inferred using the GONG 
data is shown by the continuous line, while the dotted lines indicate
the $1\sigma$ error limits.}
\label{rmu}
\end{figure}

\subsection{Cross-section for pp Reaction}

With the help of the inverted density, temperature and hydrogen
abundance profiles, it is possible to compute the total energy
generated by nuclear reactions, and this should be compared with
the observed solar luminosity, $L_\odot=3.846\times10^{33}$ ergs/sec.
As emphasized by Antia \& Chitre (\cite{ac98}) there is an ($2\sigma$)
uncertainty of about 3\% in computing the luminosity of seismic models.
This arises from possible
errors in primary inversion, solar radius, equation of state, nuclear
reaction rates for other reactions.
The uncertainty arising from errors
in $Z$ profiles is much larger and hence in this work we use seismic models
with homogeneous $Z$ profile, covering a wide range of $Z$ values.
For each central value of $Z$ we estimate the range of
cross-section of pp nuclear reaction, which reproduces the luminosity
to within 3\% of the observed value. The results are shown in Fig.~2,
which delineates the region in $Z_c$--$S_{11}$ plane that is
consistent with helioseismic and luminosity constraints.

\begin{figure}
\resizebox{\figwidth}{!}{\includegraphics{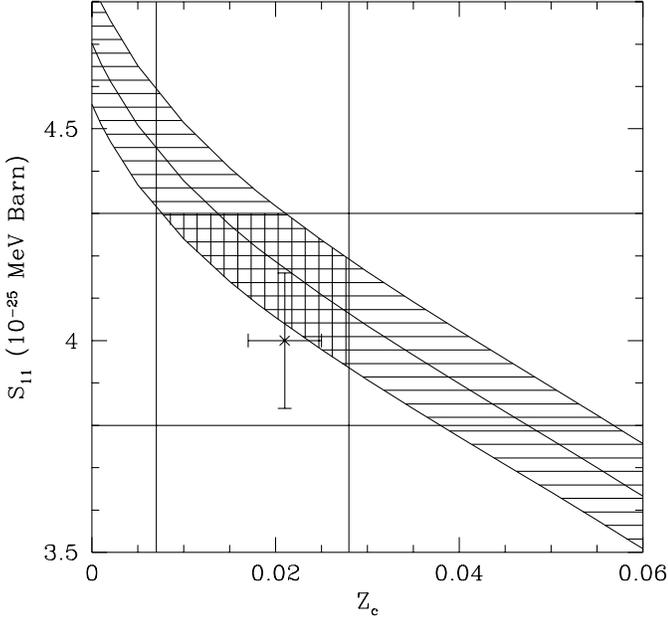}}
\caption{
The region in $Z_c$--$S_{11}$ plane that is consistent with helioseismic
data is marked by horizontal shading.
The central line defines the values where
the seismic model matches the observed solar luminosity.
The point
with $2\sigma$ error bars shows the current best estimates for $Z_c$ and
$S_{11}$. The vertical lines denote the limits on central $Z$ values
obtained by Fukugita \& Hata~(\cite{fuk98}) and the horizontal lines
mark the limits on $S_{11}$ as
obtained by various calculations so far. The region with
vertical shading indicates the area that is consistent with all data.
}
\end{figure}

It can be seen that current best estimates for $Z_c$ and $S_{11}$
(Bahcall, Basu \& Pinsonneault \cite{bp98}) are only
marginally consistent with helioseismic constraints and probably need to
be increased slightly. This figure also shows the limits on the values of
$Z_c$ obtained by Fukugita \& Hata~(\cite{fuk98}) as well as the range of
$S_{11}$ as inferred from various theoretical calculations so far
(Bahcall \& Pinsonneault \cite{bp95}; Turck-Chi\'eze \& Lopes \cite{tc93}).
One therefore,
expects that the values of $Z_c$ and $S_{11}$ should fall within the
region with vertical shading in Fig.~2.

\begin{figure}
\resizebox{\figwidth}{!}{\includegraphics{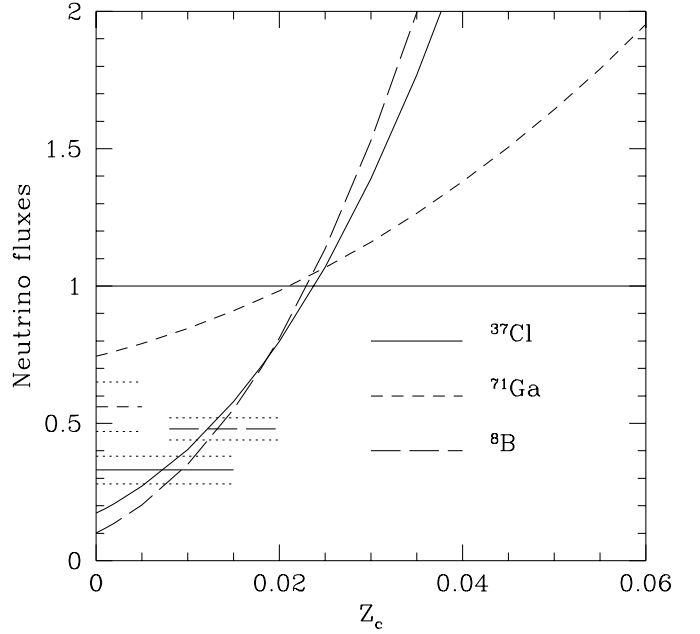}}
\caption{
The neutrino fluxes scaled in terms of those in standard solar
model (Bahcall et al.~\cite{bp98})  are displayed
as a function of heavy element abundance in the solar
core, for the seismic model with the correct observed luminosity.
For each neutrino experiment, the 
horizontal lines mark the observed value with dotted
lines denoting the $2\sigma$ error limits.
The error bars on computed values is not shown for clarity.
These error estimates can be found in Table~1 of Antia \& Chitre~(1998).
}
\end{figure}

The neutrino fluxes in seismic models with the correct luminosity
(for the value of $S_{11}$ corresponding to the central line in Fig.~2)
as a function of $Z_c$ are shown in Fig.~3. It can be seen that the neutrino
flux in $^{71}$Ga detector is never as low as the observed value, while
the $^8$B neutrino flux and the neutrino flux in $^{37}$Cl are within
observed limits, although for disjoint values of $Z_c$. Thus, a variation
of $Z_c$ values does not yield
neutrino fluxes that are simultaneously consistent
with any two of the three solar neutrino experiments.
Similar conclusions were reached  from more general considerations by
Hata, Bludman \& Langacker~(\cite{hat94}),
Heeger \& Robertson~(\cite{hee96}), Bahcall~(\cite{bah96}),
Castellani et al.~(\cite{cas97}), Antia \& Chitre~(\cite{ac97}).

\subsection{Determination of $X$ and $Z$ profiles}

It is clear that $Z$ profile is the major source of uncertainty
in helioseismic
constraint on the pp nuclear reaction cross-section.
We, therefore, explore the possibility of determining
the $Z$ profile in addition to the
$T,X$ profiles using the equations of thermal equilibrium, along with
the sound speed, density and pressure profiles. This would require
a determination of two of the three unknowns
$T,X,Z$, with the two constraints obtained from primary inversions,
namely, $p(T,\rho,X,Z)$ and $c(T,\rho,X,Z)$. 
We can thus write
\begin{eqnarray}
\noalign{\medskip}
{\delta c\over c}&=&\left(\partial\ln c\over\partial\ln\rho\right)_{T,X,Z}
{\delta\rho\over\rho}+
\left(\partial\ln c\over\partial\ln T\right)_{\rho,X,Z}{\delta T\over T}
\nonumber\\
&&\qquad +\left(\partial\ln c\over\partial X\right)_{\rho,T,Z}\delta X+
\left(\partial\ln c\over\partial Z\right)_{\rho,T,X}\delta Z, \label{dc}\\
\noalign{\bigskip}
{\delta p\over p}&=&\left(\partial\ln p\over\partial\ln\rho\right)_{T,X,Z}
{\delta\rho\over \rho}+
\left(\partial\ln p\over\partial\ln T\right)_{\rho,X,Z}{\delta T\over T}
\nonumber\\
&&\qquad +\left(\partial\ln p\over\partial X\right)_{\rho,T,Z}\delta X+
\left(\partial\ln p\over\partial Z\right)_{\rho,T,X}\delta Z.\label{dp}\\
\nonumber
\end{eqnarray}

\noindent
Since $\rho$ is known independently, we ignore the variation in $\rho$
and consider only $T,X,Z$. Now for a fully ionized nonrelativistic perfect
gas, it is well known that 
\begin{eqnarray}
\noalign{\medskip}
2\left(\partial\ln c\over\partial\ln T\right)_{\rho,X,Z}&=&
\left(\partial\ln p\over\partial\ln T\right)_{\rho,X,Z}=1,\\
\noalign{\bigskip}
2\left(\partial\ln c\over\partial X\right)_{\rho,T,Z}&=&
\left(\partial\ln p\over\partial X\right)_{\rho,T,Z}
\approx{5\over5X+3-Z},\\
\noalign{\bigskip}
2\left(\partial\ln c\over\partial Z\right)_{\rho,T,X}&=&
\left(\partial\ln p\over\partial Z\right)_{\rho,T,X}
\approx-{1\over5X+3-Z},\\
\nonumber
\end{eqnarray}
It is clearly not possible to determine any two of these three
quantities $T,X,Z$, from $c$ and $p$, since if the $\rho$ variations
are ignored,
we always have $2\delta c/c=\delta p/p$, and these constraints are
not independent. Thus, we need to check if the actual equation of state
used in solar model computations allows these quantities to be
independent.
Another basic problem in trying to determine $Z$ using Eqs.~(3--4) is
that in general we would expect $|\delta Z|<<|\delta X|$, while
the derivatives w.r.t. $Z$ are smaller than those w.r.t. $X$ and
hence we would expect the $\delta Z$ term to be much smaller than the
$\delta X$ term, making it difficult to determine $Z$ using these
equations. Thus we can only hope to use these equations to determine
$T$ and $X$, while $Z$ can be determined from equations of thermal
equilibrium through the opacity, which depends sensitively on $Z$.

\begin{figure}
\resizebox{\figwidth}{!}{\includegraphics{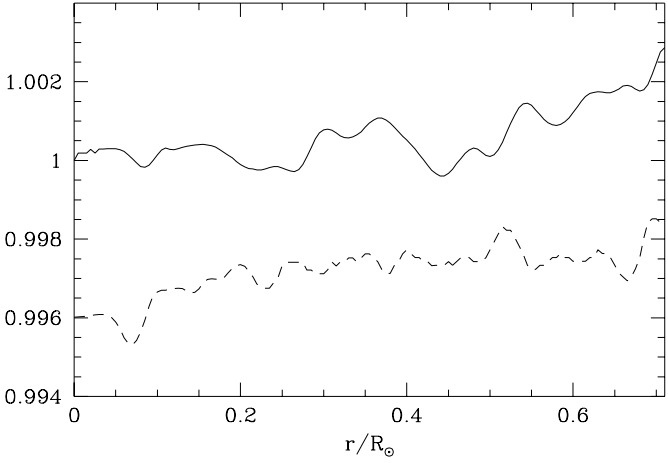}}
\caption{
The continuous line shows the ratio ${\partial \ln c^2\over\partial X}
/{\partial \ln p\over\partial X}$ for a solar model, while
the dashed line displays the ratio ${\partial \ln c^2\over\partial \ln T}
/{\partial \ln p\over\partial \ln T}$.}
\end{figure}

Fig.~4 shows the ratio of partial derivatives for $c^2$ and
$p$, as a function of $r$ in a solar model
and it is clear that these derivatives are almost equal. The wiggles in the
curve are probably due to errors in estimating these derivatives
and it is clear that the departure of the ratio
from unity is comparable to these errors, particularly, for the
derivatives with respect to $X$. Thus, for the 
solar case these two constraints are not independent and it is
demonstrably not possible
to get any additional information by using the pressure profile.
Any attempt to do so will only yield arbitrary results magnifying the errors
arising from those in the equation of state and primary inversions.

In order to estimate the extent of error magnification we can try to
compute the ratio
\begin{equation}
R_{T,X}={\left(\partial\ln c\over\partial\ln T\right)_{\rho,X,Z}
\left(\partial\ln p\over\partial X\right)_{\rho,T,Z}
\over \left(\partial\ln c\over\partial\ln T\right)_{\rho,X,Z}
\left(\partial\ln p\over\partial X\right)_{\rho,T,Z}-
\left(\partial\ln p\over\partial\ln T\right)_{\rho,X,Z}
\left(\partial\ln c\over\partial X\right)_{\rho,T,Z}},
\end{equation}
and similar ratios between derivatives with respect to $(T,Z)$ or
$(X,Z)$. It turns out that all these quantities are greater than 200 over
the entire solar model. Thus all errors will be magnified by a factor
of at least 200, if we attempt to determine the $Z$ profile, in addition to
$T,X$ profiles.

Even if we do not impose the additional constraint arising from pressure,
we can calculate the pressure profile using the OPAL equation of
state from the
inferred $T,\rho,X$ and assumed $Z$ profiles. As mentioned earlier,
we also apply the relativistic corrections (Elliot \& Kosovichev
\cite{ell98}) to the equation of state.
This $p$-profile can be
compared with that inferred from primary inversions using the equation
of hydrostatic equilibrium and Fig.~5 shows the results. It is clear that
even without applying the additional constraint from $p(T,\rho,X,Z)$ the
resulting profile comes out to be very close to the
``independently'' inferred profile,
well within the $1\sigma$ error limits.
Moreover, the inferred profile
is rather insensitive to $Z$ and hence effecting a change in $Z$ is
unlikely to produce the profiles that will 
match the primary inversion exactly.
It is, therefore, evident that the pressure profile does not provide an
independent constraint. There are only two independent constraints
(e.g., $c,\rho$) that can be calculated from the primary inversions
and it becomes well nigh impossible
to determine $Z$ profile in addition to the $T,X$ profiles.

\begin{figure}
\resizebox{\figwidth}{!}{\includegraphics{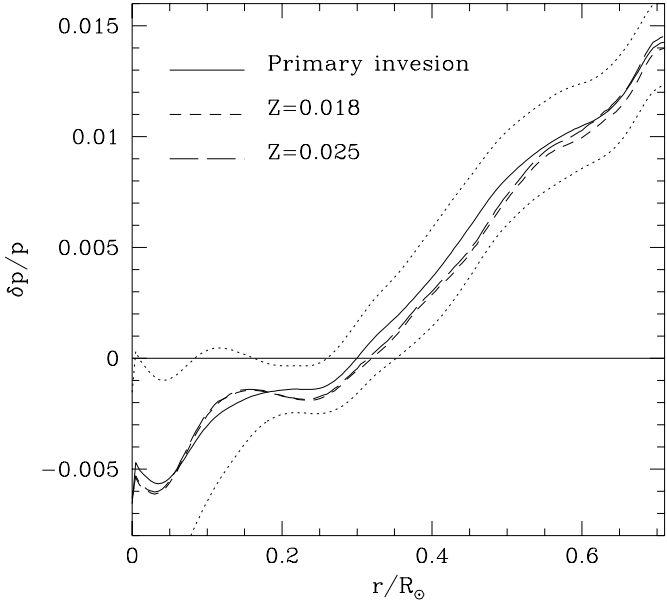}}
\caption{
The relative difference in pressure between the Sun and 
Model~S (\jcd\ et al.~\cite{jcd96}) as inferred by primary inversion and by
secondary inversion using the OPAL equation of state and labelled value
of $Z$ at the surface. The dotted  lines are $1\sigma$ errors in
primary inversion.}
\end{figure}

\subsection{Computation of $Z$ profile}

We have stressed earlier that it is not feasible to determine
both $X$ and $Z$ profiles,
in addition to the temperature, from equations of thermal equilibrium and
primary inversions. However, we can reverse the process and determine the
$Z$ profile instead of the $X$ profile, using these equations.
We, therefore, prescribe an $X$ profile from some solar model and seek
to determine the $Z$ profile using the equations described earlier.
In this case the equation of state $c=c(T,\rho,X,Z)$ is used to
determine $T$ and then using Eqs.~(1--2) we can determine $L_r$ and
$\kappa$. From the opacity $\kappa$ we can determine the required
value of $Z$ using the OPAL opacity tables. Thus in this process we
would also get an estimate of opacity variations required to
make the solar model consistent with helioseismic data. This is similar
to what has, indeed, been done by
Tripathy \& \jcd~(\cite{tri98}) except for the fact that they have
used only the inverted sound speed profile, while we
constrain, in addition, the density profile.

\begin{figure}
\resizebox{\figwidth}{!}{\includegraphics{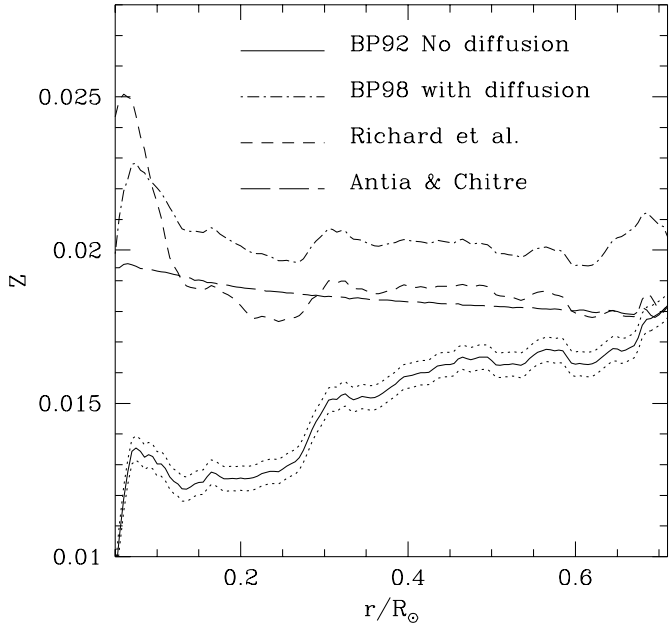}}
\caption{
The $Z$ profiles inferred using a prescribed profile for
$X$ from different solar models as labelled in the figure.
For clarity, only for one profile the error estimates are shown with
dotted lines indicating $1\sigma$ error limits.}
\end{figure}

The resulting $Z$ profiles from our calculations are shown in Fig.~6.
This figure displays the results using an $X$ profile from a model without
diffusion (Bahcall \& Pinsonneault \cite{bp92}) and some models with diffusion
(Bahcall et al.~\cite{bp98}; Richard et al.~\cite{ric96}).
From Fig.~6 it is clear that for an $X$ profile from a solar model without
diffusion, the required change in $Z$ or opacities is rather large, thus
supporting other evidence for diffusion of helium below the solar
convection zone. The long-dashed line in Fig.~6 has been
obtained using the $X$ profile inferred by Antia \& Chitre~(\cite{ac98})
with the $Z$ profile from Richard et al.~(\cite{ric96}).
The $Z$ profile is evidently reproduced, demonstrating
the consistency of the calculations.
It may be noted that the error limits displayed in this figure denote the
statistical error resulting from uncertainties in observed frequencies
and do not include systematic errors arising from other sources.
Possible errors in opacity tables may introduce
much larger uncertainties in the inferred $Z$ profile. But it is difficult
to estimate these errors and hence we have not included them in our
analysis. The only purpose of this exercise is to estimate the extent of
opacity (or $Z$) modifications required to get a solar model that is
consistent with helioseismic constraints. Of course, this does not
give us an estimate of actual error in opacity calculations as there
could be other uncertainties in solar models which have not been
addressed.

\section{Discussion and Conclusions}

Using the primary inversions for $c,\rho$, it is possible to infer
the $T,X$ profiles in solar interior, provided $Z$ profile is known.
The resulting seismic models have the correct solar
luminosity only when the heavy element abundance $Z_c$ in the solar core and
the cross-section for pp nuclear reaction rate are within the
shaded region shown in Fig.~2. It appears that the
currently accepted values of $Z_c$
or $S_{11}$ need to be increased marginally to make them consistent
with helioseismic constraints.

It is not possible to uniquely determine all three quantities
$T,X,Z$ using equations of thermal equilibrium along with results
from primary inversions, as there are only two independent constraints
that emerge from primary inversions. Incorporation of the pressure
profile as an additional input from primary inversions does not yield an
independent constraint for determining  $Z$, in addition to
$T$ and $X$.
However, it may be possible to determine the $Z$ profile
using equations of thermal equilibrium, provided the $X$ profile is
independently prescribed. This gives an estimate
of variation in opacity required to match the helioseismic data.
From these results it is clear that $X$ profile for solar models
without diffusion of helium is not consistent with helioseismic
data, unless opacity (or $Z$) is reduced by a large amount.

\begin{acknowledgements}
This work utilizes data obtained by the Global Oscillation Network
Group (GONG) project, managed by the National Solar Observatory, a
Division of the National Optical Astronomy Observatories, which is 
operated by AURA, Inc. under a cooperative agreement with the 
National Science Foundation.
The data were acquired by instruments operated by the Big
Bear Solar Observatory, High Altitude Observatory,
Learmonth Solar Observatory, Udaipur Solar Observatory,
Instituto de Astrofisico de Canarias, and Cerro Tololo
Interamerican Observatory.
\end{acknowledgements}


\begin{thebibliography}{}

\bibitem[1998]{fusion} Adelberger E. C., Austin S. M.,
Bahcall J. N. et al.\ 1998, Rev.\ Mod.\ Phys.\ 70, 1265

\bibitem[1996]{a96}  Antia H. M., 1996, A\&A 307, 609

\bibitem[1995]{ac95} Antia H. M., Chitre S. M., 1995, ApJ 442, 434

\bibitem[1997]{ac97} Antia H. M., Chitre S. M., 1997, MNRAS 289, L1

\bibitem[1998]{ac98} Antia H. M., Chitre S. M., 1998, A\&A 339, 239

\bibitem[1996]{bah96} Bahcall J. N., 1996, ApJ 467, 475

\bibitem[1992]{bp92} Bahcall J. N., Pinsonneault M. H., 1992,
Rev.\ Mod.\ Phys.\ 64, 885 

\bibitem[1995]{bp95} Bahcall J. N., Pinsonneault M. H., 1995,
Rev.\ Mod.\ Phys.\ 67, 781

\bibitem[1998]{bp98} Bahcall J. N., Basu S., Pinsonneault M. H., 1998,
Phys.\ Let.\ B 433, 128

\bibitem[1997]{cas97} Castellani V., Degl'Innocenti S., Fiorentini G.,
Lissia M., Ricci B., 1997, Phys. Rep. 281, 309

\bibitem[1996]{jcd96} \jcd\ J., D\"appen W., et al.,~1996, Sci.~272, 1286

\bibitem[1998]{inn98} Degl'Innocenti S., Fiorentini G., Ricci B.,
1998, Phys. Lett. B416, 365

\bibitem[1998]{ell98} Elliott, J. R., Kosovichev, A. G., 1998, 
ApJ 500, L199

\bibitem[1998]{fuk98} Fukugita M., Hata N., 1998, ApJ 499, 513

\bibitem[1988]{dog88} Gough D. O., Kosovichev A. G., 1988, in
Seismology of the Sun and Sun-like Stars, eds.\ V. Domingo \&
E. J. Rolfe, ESA Publ. SP-286, p.195.

\bibitem[1994]{hat94} Hata N., Bludman S., Langacker P., 1994,
Phys.\ Rev.\ D49, 3622

\bibitem[1996]{hee96} Heeger K. M., Robertson R. G. H., 1996,
Phys.\ Rev.\ Lett.\ 77, 3720

\bibitem[1996]{hil96} Hill F.,  Stark P. B., Stebbins R. T., et al.~1996,
Sci.\ 272, 1292

\bibitem[1996]{igl96} Iglesias C. A., Rogers F. J., 1996, ApJ 464, 943

\bibitem[1996]{kos96} Kosovichev A. G., 1996, Bull.\ Astron.\ Soc.\ India
24, 355

\bibitem[1996]{ric96} Richard O., Vauclair S., Charbonnel C.,
Dziembowski W. A., 1996, A\&A 312, 1000

\bibitem[1996]{rog96} Rogers F. J., Swenson F. J., Iglesias C. A.,
1996, ApJ 456, 902

\bibitem[1996]{rox96} Roxburgh I. W., 1996, Bull.\ Astron.\ Soc.\ India
24, 89

\bibitem[1998]{bon99} Schlattl H., Bonanno A., Patern\'o L.,
1998, Astro-ph/9902354

\bibitem[1996]{st96} Shibahashi H., Takata M., 1996, PASJ 48, 377 

\bibitem[1998]{shi98} Shibahashi H., Hiremath K. M., Takata M., 1997,
in Proc. IAU Symp.\ No.\ 185: New Eyes to See Inside the Sun and Stars,
eds., F.-L. Deubner, J. Christensen-Dalsgaard and D. W. Kurtz
(Kluwer: Dordrecht) p81

\bibitem[1998]{tak98} Takata M., Shibahashi H.,  1998,
ApJ 504, 1035

\bibitem[1998]{tri98} Tripathy S. C., \jcd\ J., 1998, A\&A 337, 579

\bibitem[1993]{tc93} Turck-Chi\'eze S.,  Lopes I., 1993, ApJ 408, 347

\end{thebibliography}
\end{document}